\documentclass[3p,times,twocolumn]{elsarticle}
\usepackage{color}
\usepackage{amssymb,amsmath}
\usepackage{url}
\usepackage{multirow}
\journal{}

\begin{document}

\begin{frontmatter}

\title{Light Yield in DarkSide-10: a Prototype Two-phase Argon TPC for Dark Matter Searches}

\author[fnal]{T.~Alexander}
\author[augustana]{D.~Alton}
\author[ucla]{K.~Arisaka}
\author[princeton]{H.O.~Back}
\author[ucla]{P.~Beltrame}
\author[princeton_chem]{J.~Benziger}
\author[lngs]{G.~Bonfini}
\author[mi]{A.~Brigatti}
\author[princeton]{J.~Brodsky}
\author[umass]{L.~Cadonati}
% ...
\address[augustana]{Physics and Astronomy Department, Augustana College, Sioux Falls, SD 57197, USA}
% ...
\address[fnal]{Fermi National Accelerator Laboratory, Batavia, IL 60510, USA}
% ...
\address[lngs]{Laboratori Nazionali del Gran Sasso, SS 17 bis Km 18+910, Assergi (AQ) 67010, Italy}
% ...
\address[princeton_chem]{Chemical Engineering Department, Princeton University, Princeton, NJ 08544, USA}
\address[princeton]{Physics Department, Princeton University, Princeton, NJ 08544, USA}
% ...
\address[mi]{Physics Department, Universit\`a degli Studi and INFN, Milano 20133, Italy}
% ...
\address[ucla]{Physics and Astronomy Department, University of California, Los Angeles, CA 90095, USA}
% ...
\address[umass]{Physics Department, University of Massachusetts, Amherst, MA 01003, USA}
% ...
\cortext[cor1]{Corresponding author}

\begin{abstract}
As part of the DarkSide program of direct dark matter searches using two-phase argon TPCs, a prototype
detector with an active volume containing 10 kg of liquid argon, DarkSide-10, was built and operated underground
in the Gran Sasso National Laboratory in Italy.  A critically important parameter for such devices is the scintillation light yield,
as photon statistics limits the rejection of electron-recoil backgrounds by pulse shape discrimination.  We have measured the light yield
of DarkSide-10 using the readily-identifiable full-absorption peaks from gamma ray sources combined with single-photoelectron calibrations 
using low-occupancy laser pulses.  For gamma lines of energies in the range 122-1275 keV, we get light yields averaging 
8.887$\pm$0.003(stat)$\pm$0.444(sys)~p.e./keV$_{ee}$.  With additional purification, the light yield measured at 511 keV increased to 9.142$\pm$0.006(stat)~p.e./keV$_{ee}$.
\end{abstract}


\end{frontmatter}

\section {Introduction}
\indent
Particle detectors based on liquid argon, first developed in the 1970's for calorimetry and tracking~\cite{Rubbia:117852}, have recently become recognized as an extremely attractive technology for the direct detection of dark matter~\cite{darkside, Boulay:2008bc, Benetti2008495, Rubbia:2006ar}. This was in large part due to the strong pulse shape discrimination (PSD) possible with liquid argon scintillation~\cite{Boulay2006179}.  Particles interacting in argon induce excitation and ionization of the medium, leading to the emission of scintillation light whose time structure is strongly correlated with the nature of the interaction~\cite{Kubota1978561}. This provides a way to detect rare nuclear recoil events, possibly induced by Weakly Interacting Massive Particles (WIMPs), a well-motivated galactic dark matter candidate.   Such events must be identified in the presence of an overwhelming background of low-ionization-density electron-induced events from background $\gamma$ and $\beta$ radioactivity.  With sufficient photon statistics, PSD can allow discrimination of nuclear recoil events from electron-induced background events at better than $10^{-8}$~\cite{Benetti2008495, Boulay2006179, Boulay:2012hq, PhysRevC.78.035801}.  If realized in a low-background detector, this offers a natural route to a sensitive dark matter search. 

Because the efficiency of the PSD is so strongly dependent on the number of detected scintillation photons, much  of the recent R\&D activity in the field has been aimed at improving the light collection of liquid-argon-based detectors.

The DarkSide project is a direct detection dark matter search at Laboratori Nazionali del Gran Sasso (LNGS). It will be based on the use of a two-phase Time Projection Chamber (TPC) filled with argon from underground sources, strongly depleted in the naturally occurring $^{39}$Ar radioisotope~\cite{AcostaKane200846}.  The detector will be coupled with a borated liquid scintillator neutron veto, resulting in a detector capable of achieving background-free operation in an extended run~\cite{darkside, Wright201118}. 

DarkSide-50, currently being commissioned with an active mass of 50 kg, will be the first physics-capable detector in the DarkSide program. In order to develop and optimize the two-phase argon time projection technology, a dedicated 10-kg prototype (DS-10) was built at Princeton University. After seven months of running at Princeton, the DS-10 detector was deployed at LNGS during the spring of 2011. We report measurements of its light yield, performed during the first underground data campaign, July-December, 2011.
\section{Scintillation Light}
\indent 
Scintillation in liquid argon results from the radiative decays of excited molecular dimers, Ar$_2^*$, formed  after the passage of charged particles. An excited dimer may be formed in either a singlet or a triplet state, which have very different radiative decay lifetimes, $\sim\!\! 7$ ns for the singlet and $\sim\!\! 1.6$ $\mu$s for the triplet~\cite{Kubota1978561}. The relative population of the fast (singlet) and slow (triplet) components is strongly correlated with the ionization density and hence the nature of the primary ionizing particle and the deposited energy~\cite{PhysRevB.27.5279}. The typical fraction of the scintillation light in the fast component is $\sim\!\! 0.7$ for nuclear recoil events, which are heavily-ionizing, and  $\sim\!\! 0.3$ for electron-mediated events -- this is the basis of pulse shape discrimination.

The rejection power achievable by PSD is strongly affected by the amount of light detected.  This is due to the  growing statistical precision  with which the fast and slow component populations can be determined  for any event type as the total number of detected photons increases. Consequently, the overall light collection and detection efficiency becomes a crucial figure of merit for the performance of these detectors.  With the common use of photomultiplier tubes (PMTs) as light sensors,  the overall  light yield is often expressed in terms of the number of detected photoelectrons (p.e.) per keV of energy deposited in the argon. This number, which in general depends on the nature of the ionizing particle and on the applied electric field, is usually quoted for electron recoil events at null field, and expressed in units of p.e./keV$_{ee}$ (for ``electron equivalent" energy). In absolute terms, the detector light yield can be compared to the raw photon yield in liquid argon, $\sim\!\! 40$ scintillation photons per keV$_{ee}$ deposited~\cite{Doke1988291}. This scintillation light is peaked in the UV at 128 nm.

We measure the light yield of the DS-10 detector by studying the scintillation  spectra from radioactive $\gamma$ sources deployed outside the cryostat. By normalizing the integrated signal on each PMT to the average integral corresponding to a single photoelectron, the light yield of the detector can be estimated from the spectral features of the $\gamma$ sources.  

\section{The DS-10 Detector}
\begin{figure}[t]
\begin{center}
\includegraphics[width = \linewidth]{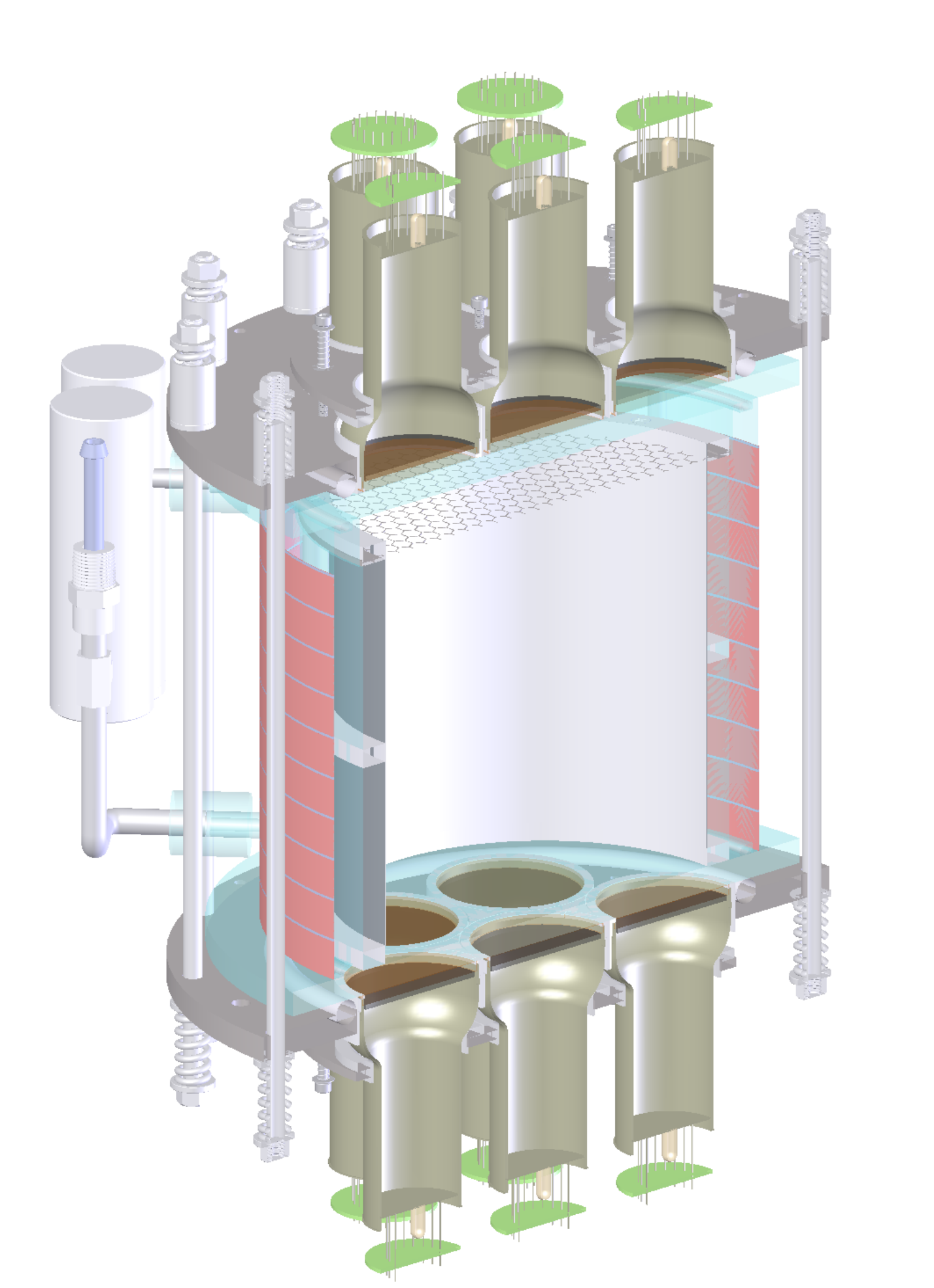}
\caption{Vertically-sectioned drawing of the DS-10 detector.}
\label{fig:ds-10}
\end{center}
\end{figure}

\indent
The DarkSide-10 detector, shown in Fig.~\ref{fig:ds-10}, is a two-phase (liquid and gas) TPC~\cite{Loer:2011bl},  incorporating several innovative design features intended to improve stability, simplicity, and performance. The device consists of an acrylic/fused silica vessel which contains the two-phase sensitive region. This ``inner vessel"  is completely immersed in a liquid argon (LAr) bath contained in a vacuum-insulated stainless steel Dewar.   The PMTs are in the outer LAr bath, viewing the active volume of LAr through the inner vessel.  

The inner vessel consists of an open-ended acrylic cylinder of 23.5~cm height, 24.1~cm inner diameter, and 1.9~cm wall thickness, sealed by PTFE-encapsulated steel-spring Creavey o-rings at the top and bottom to fused silica windows, 1.3~cm thick~\cite{creavey}. The cylinder and windows are clamped together by a cage of spring-loaded, 0.95-cm-diameter stainless steel rods. The resulting seal is sufficiently ``bubble tight" to contain the argon gas pocket required for two-phase operation.

The gas for the pocket is produced in a tube running vertically alongside the acrylic cylinder.  LAr purified in the recirculation loop (see below) enters the tube and is boiled by a resistor  operating at about one watt.  A connecting pipe delivers the gas to the top of the inner vessel. The gas-liquid interface level is passively maintained  2.0 cm below the top fused silica window by a bubbler tube that vents gas from the pocket and ends in the LAr bath at the desired height. The liquid level is continuously measured by a set of discrete Pt-1000 thermistors and a capacitive level sensor in the boiling tube.  Under normal operating conditions, including gas recirculation to the inner and outer vessels, the fluctuations in inner vessel liquid level are $<1$mm.

The active volume of the detector, 21 cm in diameter, is defined by a reflector lining the acrylic cylinder. The reflector is made of overlapping sheets of 3M Vikuiti ESR~\cite{3m:2012vk}, a multilayer plastic foil, mounted inside a PTFE frame.

To detect the 128 nm argon scintillation light, we use the wavelength shifter tetraphenyl butadiene (TPB), with a peak emission wavelength of 420 nm~\cite{Burton:73}.   The TPB fluorescence decay time is $\sim\!\! 1.8$~ns, short  compared to the 7 ns fast component of the LAr scintillation~\cite{1970RSPSA.315..163P}. TPB is deposited by vacuum evaporation onto the reflector lining the acrylic cylinder and the inner surfaces of the fused silica windows.  The stainless steel mesh separating the electron drift and extraction regions of the  TPC (described below), a few cm$^2$ of the reflector covered by $\alpha$ sources, and  small gaps at the edges of the reflectors caused by differential thermal contraction are the only non-TPB-coated surfaces seen by the UV scintillation light from the active argon volume.

Measurements in a vacuum-UV spectrophotometer suggested an optimum TPB thickness of about  200\,$\mu$g/cm$^2$, a tradeoff between high  UV-to-visible conversion efficiency and low absorption of the visible light.  The reflector and windows were coated with 230-260 and 215-230 $\mu$g/cm$^2$ of TPB, respectively.  The evaporations were performed in a large high-vacuum chamber using a Knudsen effusion cell. The typical vacuum level reached prior to the evaporation was (2-7)$\times 10^{-8}$ torr. After the evaporation the parts were kept in sealed bags filled with dry argon. During the detector assembly we took care to minimize exposure of TPB-coated surfaces to air, since degradation was noted during optical-bench testing. We accomplished this by flushing the inner vessel with argon gas throughout the assembly procedure. The fused silica windows were coated about three months before the measurements reported here began.   The reflecting foils were coated 9 months before the run and were used in the preceding 3-month run in Princeton.

Wavelength-shifted scintillation light is collected by two arrays of seven Hamamatsu high-quantum-efficiency R11065 3'' PMTs~\cite{2012JInst...7.1016A}, viewing the active volume through the top and bottom fused silica windows.   A $\sim$1 mm layer of LAr optically couples the PMTs to the windows. These fused-silica-window, metal-bulb tubes are operated at negative HV and are electrically insulated from the surrounding materials with PTFE spacers. The PMTs have Hamamatsu-reported room-temperature quantum efficiencies at 420 nm ranging from 30.4 to 35.7\%, with an average of 33.9\%. They are run at a typical gain of $4\times 10^6$ with lines terminated in 50 Ohm at both ends. To enhance the light collection, the spaces between the phototubes are filled with 1.3-cm-thick PTFE reflectors, and the small exposed areas of the stainless steel endplates are covered with 3M Vikuiti foil.  

To allow the device to be operated as a time projection chamber, the inner vessel is equipped with a set of high voltage electrodes and a distribution system to provide the necessary electric fields in the sensitive volume to drift electrons to the surface of the liquid, to extract them into the gas, and to accelerate them through the gas producing a secondary  scintillation signal proportional to the collected ionization. The electrodes fixing the potential at boundaries of the chamber are: a transparent Indium Tin Oxide (ITO) cathode on the inner surface of the bottom window, a kapton flexible printed circuit board with overlapping etched copper strips alternating between the two sides wrapped around the acrylic cylinder,  an etched stainless steel grid 5 mm below the liquid-gas surface,  and an ITO anode on the bottom surface of the top window. For TPC operation, the anode is grounded and independently-controllable voltages on the cathode and grid set the drift and extraction fields, while a chain of resistors between the copper strips creates the graded potential that keeps the drift field uniform. To shield the negatively-biased PMT photocathodes from the voltages applied to the anode and cathode, each fused silica window carries a second ITO layer on its external face.  These are maintained at approximately the average of the PMT photocathode voltages. 

To obtain the light yield measurements presented here, the TPC anode, grid, and cathode as well as the field shaping copper strips have all been kept at ground potential, giving null drift, extraction, and multiplication fields. In the rest of the paper this will be referred as the \emph{null field} configuration.  In this configuration the device operates as a pure LAr scintillation detector.   The TPC system nonetheless affects the scintillation optics. The measurements presented here were made with the gas pocket present, and the gas pocket affects the light propagation, primarily through total internal reflection in the LAr.  The grid is a 100-$\mu$m-thick stainless steel membrane etched with a hexagonal pattern of through holes 0.5 cm on a side, with an optical transparency for normally incident light of 89\%. The ITO layers are 15 nm thick, the thinest thought feasible by the vendor.  Since ITO conducts, it has a complex index of refraction that results in absorption. At 420 nm, calculations give an absorption of 2\% per layer at normal incidence, and all light must pass through at least two layers to reach the PMTs. 

A 90-W Cryomech PT90 cryocooler is connected to a cold-head inside the Dewar but outside the inner vessel. The cold head provides the cooling power needed both to cool and condense argon gas in the detector during filling and to control the liquid argon temperature during normal operation. The cold-head is instrumented with a temperature sensor and a 100-W heater which are part of a feedback loop controlled by a Lakeshore 430 temperature controller. This allows the temperature of the system to be maintained at the boiling point of liquid argon (87.8 K) with typical fluctuations less than 0.1K. 

Dissolved impurities such as nitrogen, oxygen, and water  are known to strongly affect the scintillation properties of liquid argon~\cite{2010JInst...5.5003A, 2010JInst...5.6003A}. The purity of the active argon in DS-10 is established and maintained by a number of measures.  Before the detector is cooled, the dewar is repeatedly flushed and pumped over several days at room temperature using research grade (99.999\%) argon gas and a dry turbopump, achieving a final pressure of about $6\times 10^{-5}$ mbar.   This removes adsorbed impurities from metal surfaces and reduces subsequent outgassing from the internal plastic parts.  Research grade atmospheric\footnote{Atmospheric argon is used in this prototype, as opposed to the $^{39}$Ar-depleted underground argon being extracted for the DarkSide dark matter detectors~\cite{AcostaKane200846}.} argon gas is also used for the fill. The gas is further purified by a single pass through a SAES MonoTorr PS4-MT3-R1 getter which is sized and configured  to reduce O$_2$, N$_2$, and H$_2$O impurities to sub-ppb levels~\cite{Saes:2012sa}. During operation, argon purity is maintained  by continuous gas recirculation: a metal bellows pump   forces the boil-off argon from the Dewar through the MonoTorr getter, at about 15~slpm, and send it back to the cold-head for recondensation.

\section {Data Acquisition}
\indent
The data acquisition system consists of a set of 12 bit, 250 MS/s, digitizers (CAEN 1720) which record  the signals from each of the 14 photomultiplier tubes and store them for offline analysis.  To trigger the system, the anode signal from each PMT is first amplified tenfold by a LeCroy 612A fast amplifier with two parallel outputs.   One output goes directly into the digitizer channel which runs continuously, filling a circular memory buffer.  In the digitizer, one sample at one count represents 0.0078 pC from the PMT. The other output is used to form a majority trigger. This requires a coincidence, within 100 ns, of at least 5 PMTs with signals above a threshold: the latter is set independently on each channel to about 1/3$^\text{rd}$ of the photoelectron mean amplitude. When an event satisfies the majority trigger condition, data in the 14 circular buffers representing a 35~$\mu$s time window (5~$\mu$s before the trigger and 30~$\mu$s after), is downloaded to a PC and stored on a local hard disk. The acquired window length for the null field configuration has been selected to fully contain the slow component of the scintillation light, while also including relatively large pre- and post-trigger regions to allow for baseline evaluation. 

\section{Single-Photoelectron Calibration}
\label{sec:spe}
The charge response of each PMT to a single photoelectron is evaluated using a laser calibration procedure, which was repeated frequently among the data runs analyzed here.  Light pulses of $\sim$70 ps duration at 440 nm wavelength from a diode laser are injected into the detector through an optical fiber that terminates on the bottom window of the inner vessel. Diffuse reflection from the TPB leads to a roughly uniform illumination of the 14 PMTs. The controller pulses the laser at a rate of 1000 Hz and simultaneously triggers the data acquisition system. Optical filters are placed between the laser and the fiber to adjust the intensity until the average number of photoelectrons generated on each tube in any given trigger, referred to as the average occupancy, is roughly 0.1. Unlike regular data runs, the digitization window for laser runs is only 1.5 $\mu$s long. Within this record, a 0.8 $\mu$s period before the pulse arrival time is used to define the baseline.  After subtraction of this baseline, the integral of the recorded waveform is evaluated within a fixed 92-ns window around the arrival time of the laser pulse. The resulting charge spectrum for each PMT is then fitted to a model function, allowing  the mean of the single-photoelectron charge response to be determined.

The fitting function used is
\begin {align}
F(x) = \sum_{n=0}^7 P(n;\lambda)f_n(x)
\end {align}
where $P(n;\lambda)$ is a Poisson distribution with mean $\lambda$,  representing the average occupancy, and $f_n(x)$  the $n$-photoelectron charge ($x$) response of the system.  We have modeled the $n$-photoelectron response of the system as 
\begin {align}
f_n(x) = \rho(x) \ast \psi_1^{n\ast}(x)
\label{eq:npe}
\end {align}
where $\rho$ denotes the zero photoelectron response (pedestal), $\ast$ is a convolution, and $\psi_1^{n\ast}$ is the $n$-fold convolution of the PMT  single-photoelectron response function, $\psi_1$, with itself. The function representing the pedestal, $\rho$, the integral in the absence of any photoelectrons and thus the entire $n=0$ term, is described by a Gaussian, while the PMT single-photoelectron response, $\psi_1$, is modeled by the weighted sum of a decaying exponential and a Gaussian, truncated at zero,
\begin {align}
\psi_1(x) = \begin{cases}
p_E\left(\frac{1}{x_0}e^{-x/x_0}\right)+(1-p_E)\mathrm{G}(x;x_m,\sigma) & x>0;\\
0 & x\leq 0.
\end{cases}
\label{eq:pdf}
\end {align}
The Gaussian term $\mathrm{G}(x;x_m,\sigma)$ represents the single-photoelectron response from the full dynode chain, while the exponential term accounts for incomplete dynode multiplication~\cite{Dossi2000623, 2011ITNS...58.1290D}.  
\begin{figure}[t]
\begin{center}
\includegraphics[width = \linewidth]{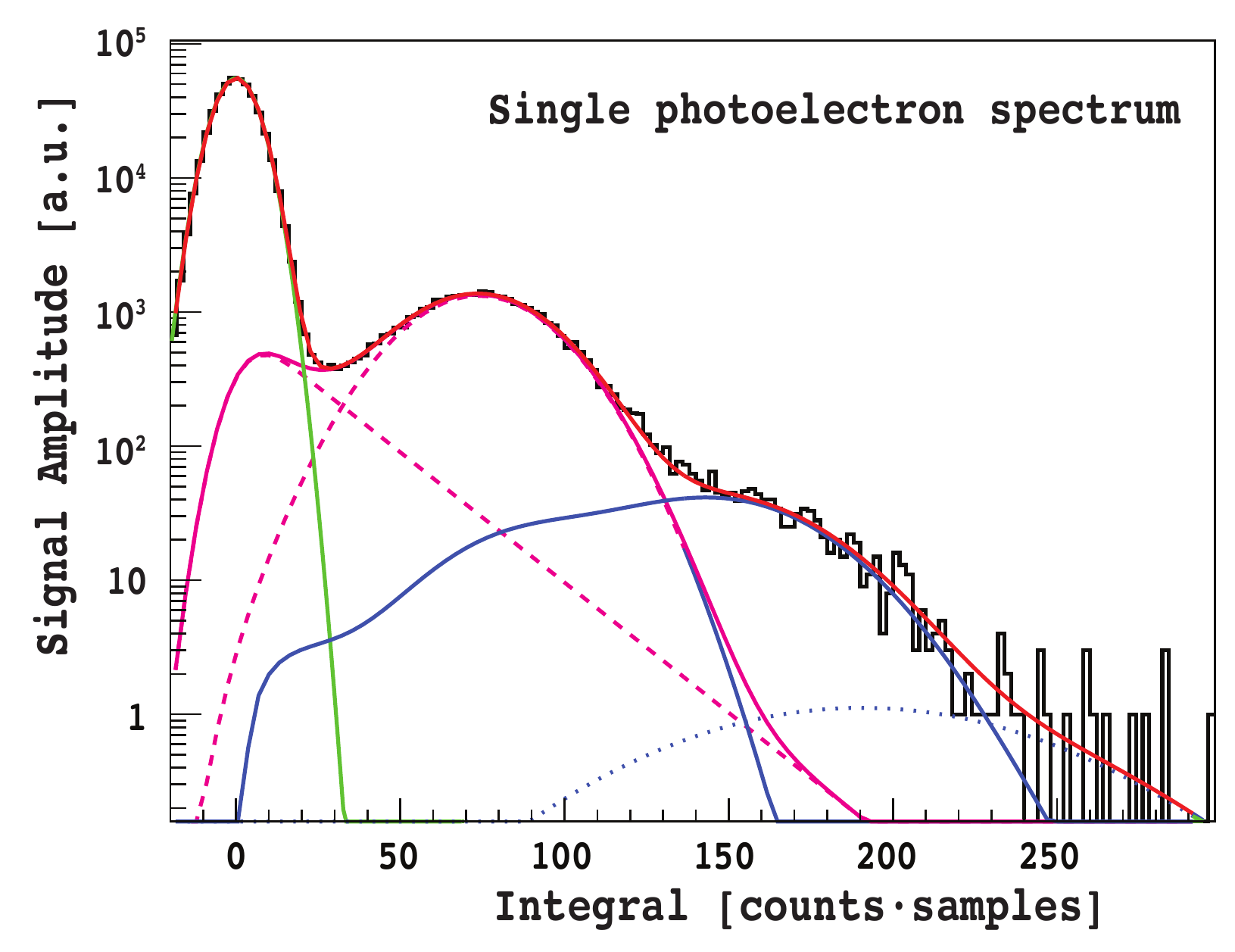}
\caption{Example of the charge response spectrum of a single PMT exposed to low-occupancy laser flashes. The horizontal axis measures charge in integrated digitizer counts (counts $\cdot$ samples), where 1~count~$\cdot$~sample corresponds to a PMT output charge of 0.0078 pC.  The colored curves represent components in the fit function used in the calibration. Green: pedestal. Dashed Magenta: Gaussian and exponential terms of the single-p.e.~model convolved with pedestal. Solid Magenta: full single-p.e.~response convolved with pedestal.  Solid Blue: 2-p.e response. Dotted Blue: $\geq 3$-p.e. response. Solid Red: Sum of all components. }
\label{fig:laser_fit}
\end{center}
\end{figure}

The fit is performed with seven free parameters: the average occupancy $\lambda$, the mean and standard deviation of the pedestal Gaussian, the mean $x_m$ and standard deviation $\sigma$ of the  single-photoelectron Gaussian, the decay constant $x_0$ of the  single-photoelectron exponential component, and the relative weight $p_E$ between the single-photoelectron Gaussian and exponential terms.  In order to simplify the computation, for $n \ge 3$ the function $\psi_1^{n\ast}$ is approximated by a Gaussian whose mean and variance are $n$ times that of the single-photoelectron response $\psi_1$. Figure~\ref{fig:laser_fit} shows a sample spectrum and fit for a laser run on a single channel. Due to the presence of the exponential term, the mean of the single-photoelectron response is, on average, 13\% lower than that of the Gaussian single-photoelectron component alone.

\section{Event Analysis}
\label{sec:source}
For each individual
channel, we determine a baseline and subtract it from the digitized waveform. Because argon scintillation pulses extend over several microseconds and eventually taper into sparse, individual photoelectrons, this needs careful treatment. The baseline is calculated by two different methods. The first, referred as the \emph{linear-baseline} method, calculates the average of the digitized samples in a gate  at least 1.0~$\mu$s wide before the trigger in the acquisition window   (where no signal is expected)  and, when possible, at the end of it, using a linear interpolation between these two values as the baseline in the region in between. If the two values differ by more than 1 ADC count, the event is rejected. The second algorithm uses a moving average, over a window of length 80~ns, to calculate a local baseline in regions where the waveform fluctuations are consistent with electronic noise. In regions where sharp excursions are found (such as under scintillation pulses, including single photoelectrons) the baseline is linearly interpolated between the most recent quiet intervals. This \emph{moving-average baseline} is intended to remove slowly varying fluctuations, such as possible low-frequency interference. 

Once the baseline has been subtracted, the waveform for each channel is divided by the corresponding single photoelectron mean, as obtained from the nearest laser calibration run. The scaled waveforms from all 14 channels are then added together to form a summed waveform which is further analyzed to identify scintillation pulses.  The scintillation-pulse finding algorithm identifies a pulse start time and end time for the scintillation signal and, once the start and end times of the  pulse are found, the integral in that interval is evaluated independently for each of the 14 scaled channels and these integrals are summed to give an estimate of the total number of photoelectrons of the scintillation event.   

\section{Response to Cobalt, Cesium, and Sodium Sources}
\begin{figure}[t!]
\begin{center}
\includegraphics[width = \linewidth]{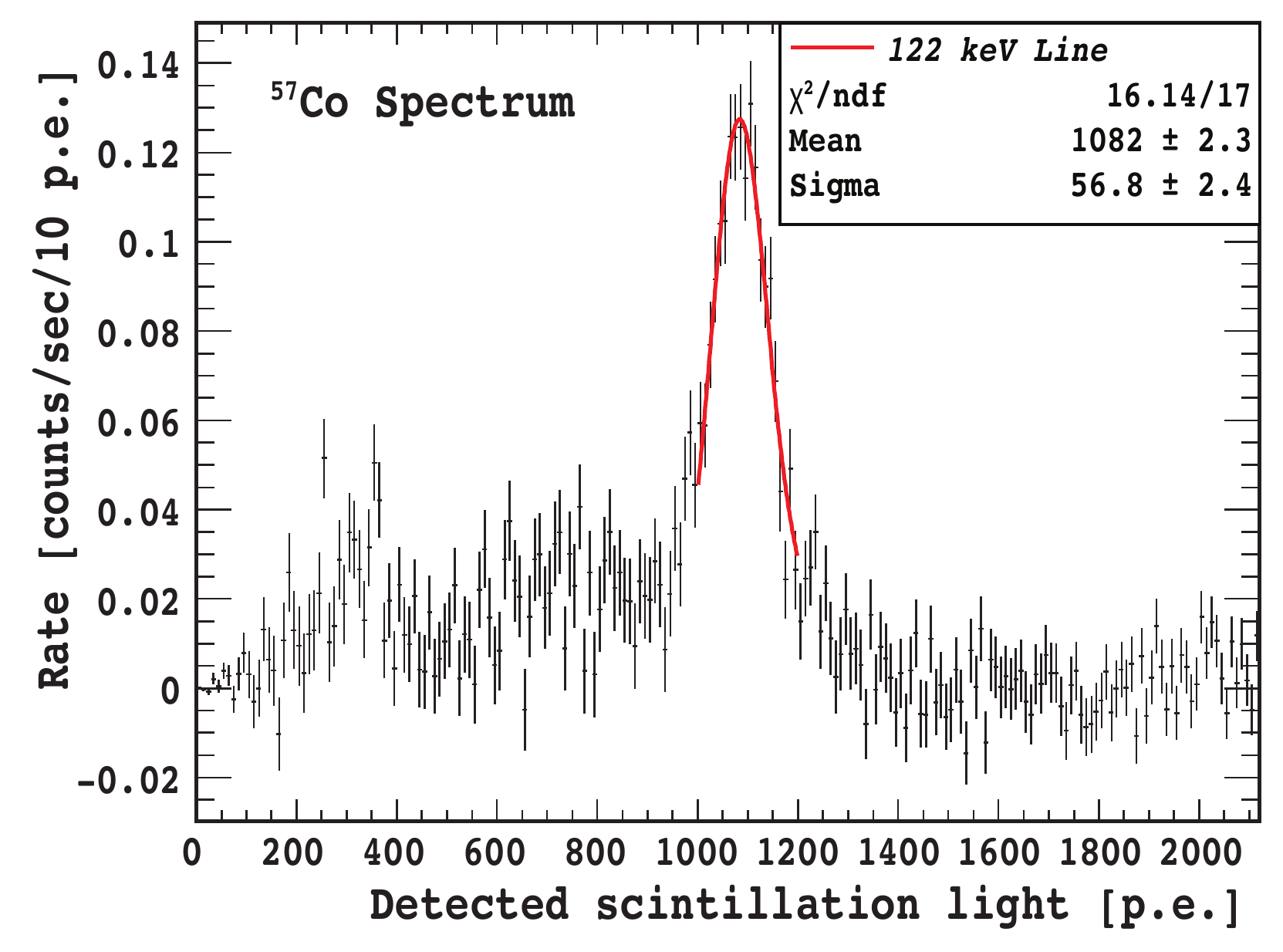}
\caption{Scintillation spectrum of $^{57}$Co collimated at the central position after subtraction of a background spectrum. The full absorption peak has been fit with two Gaussians (see text). The best-fit function is superimposed on the histogram in the energy range over which the fit was performed.}
\label{fig:spectrum_co}
\end{center}
\end{figure}

\begin{figure}[t]
\begin{center}
\includegraphics[width = \linewidth]{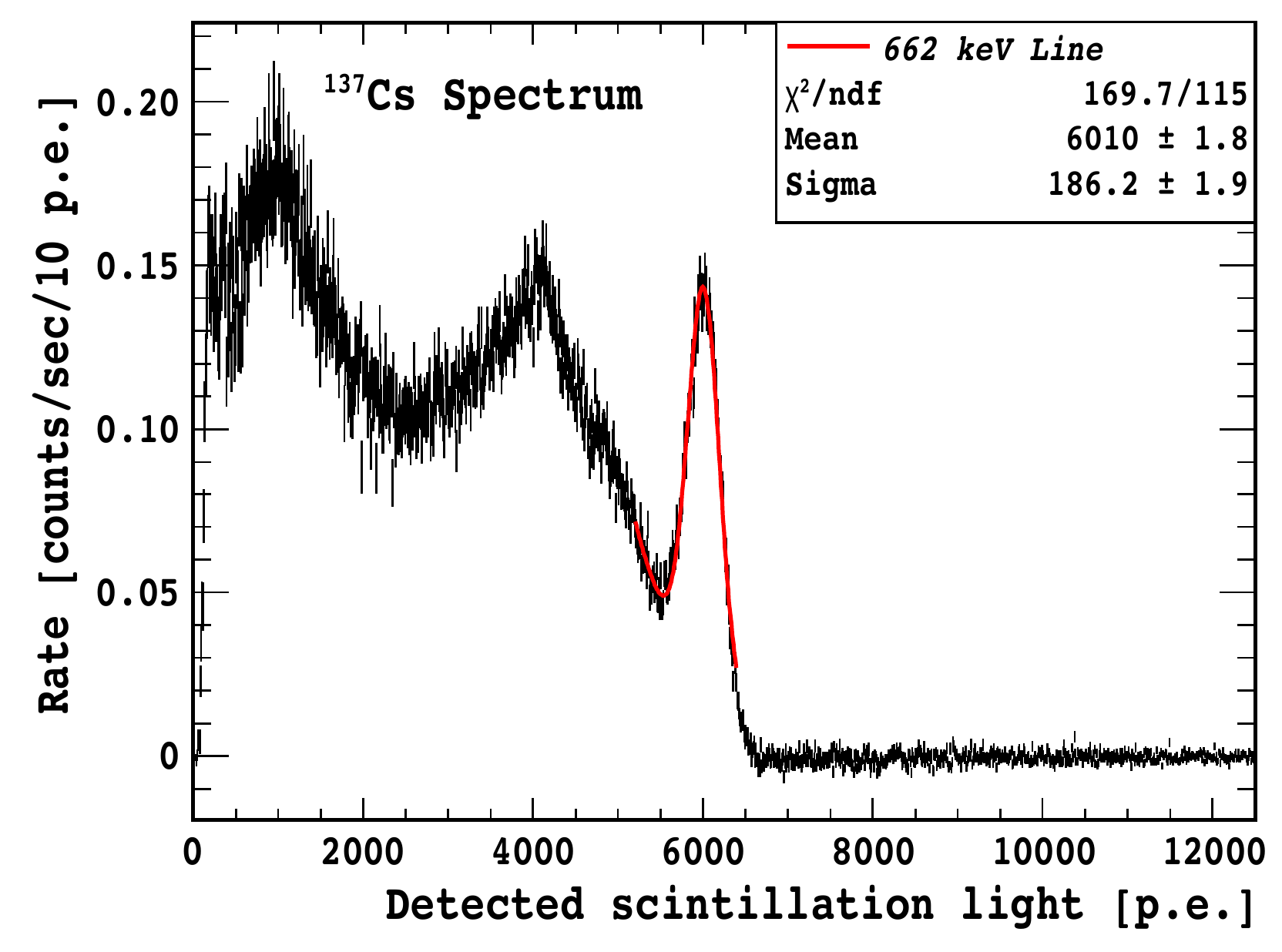}
\caption{Scintillation spectrum of $^{137}$Cs collimated at the central position after subtraction of a background spectrum. The full absorption peak has been fit with the sum of a Gaussian and a falling exponential. The best-fit function is superimposed on the histogram in the energy range over which the fit was performed.}
\label{fig:spectrum_cs}
\end{center}
\end{figure}

\begin{figure}[t]
\begin{center}
\includegraphics[width = \linewidth]{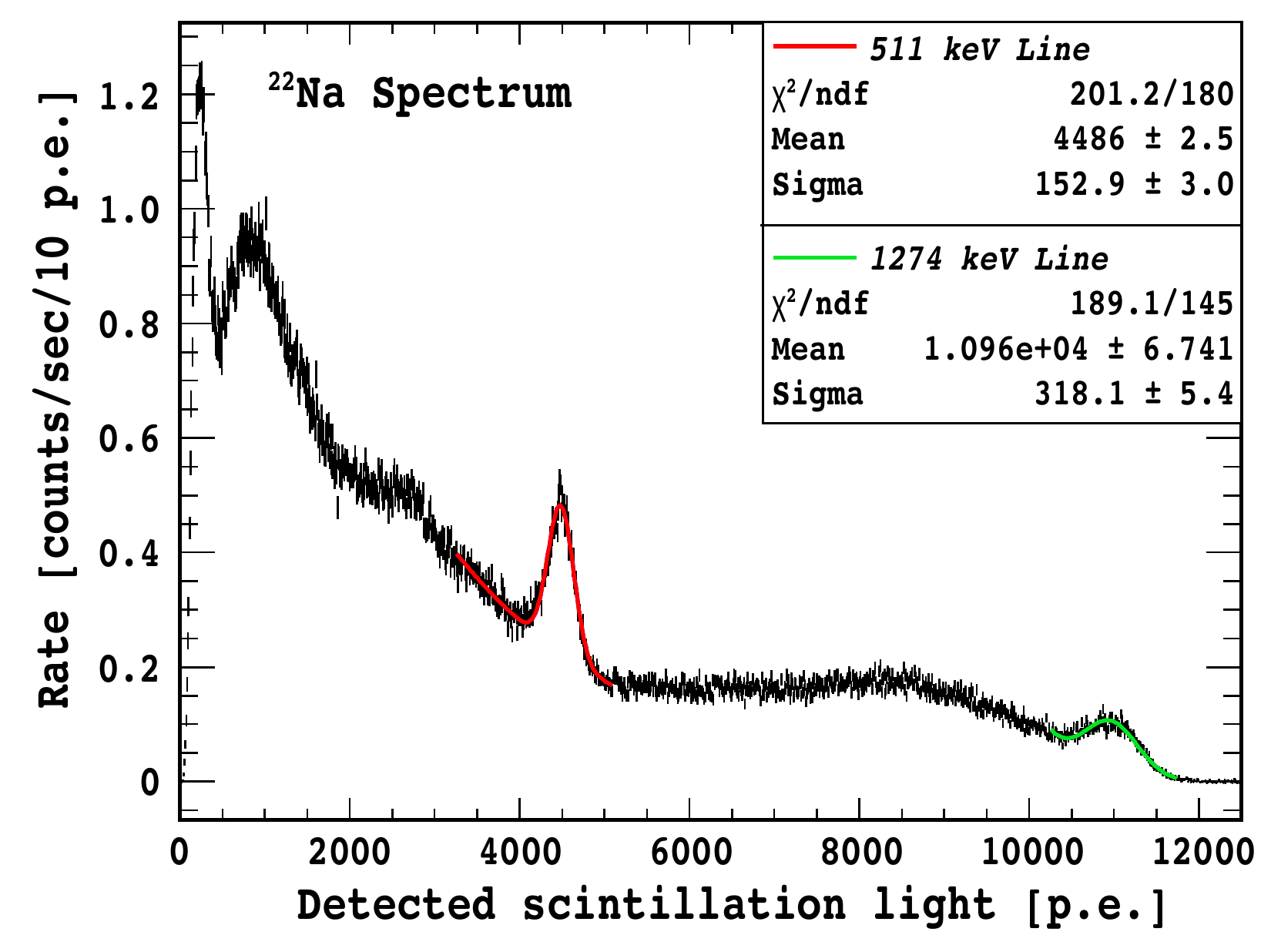}
\caption{Scintillation spectrum of $^{22}$Na collimated at the central position after subtraction of a background spectrum. The full absorption peaks at 511~keV and 1274~keV have been fitted with the sum of a Gaussian and a falling exponential. The best-fit functions are superimposed on the histogram in the energy ranges over which the fits were performed.}
\label{fig:spectrum_na}
\end{center}
\end{figure}

\begin{figure}[t]
\begin{center}
\includegraphics[width = \linewidth]{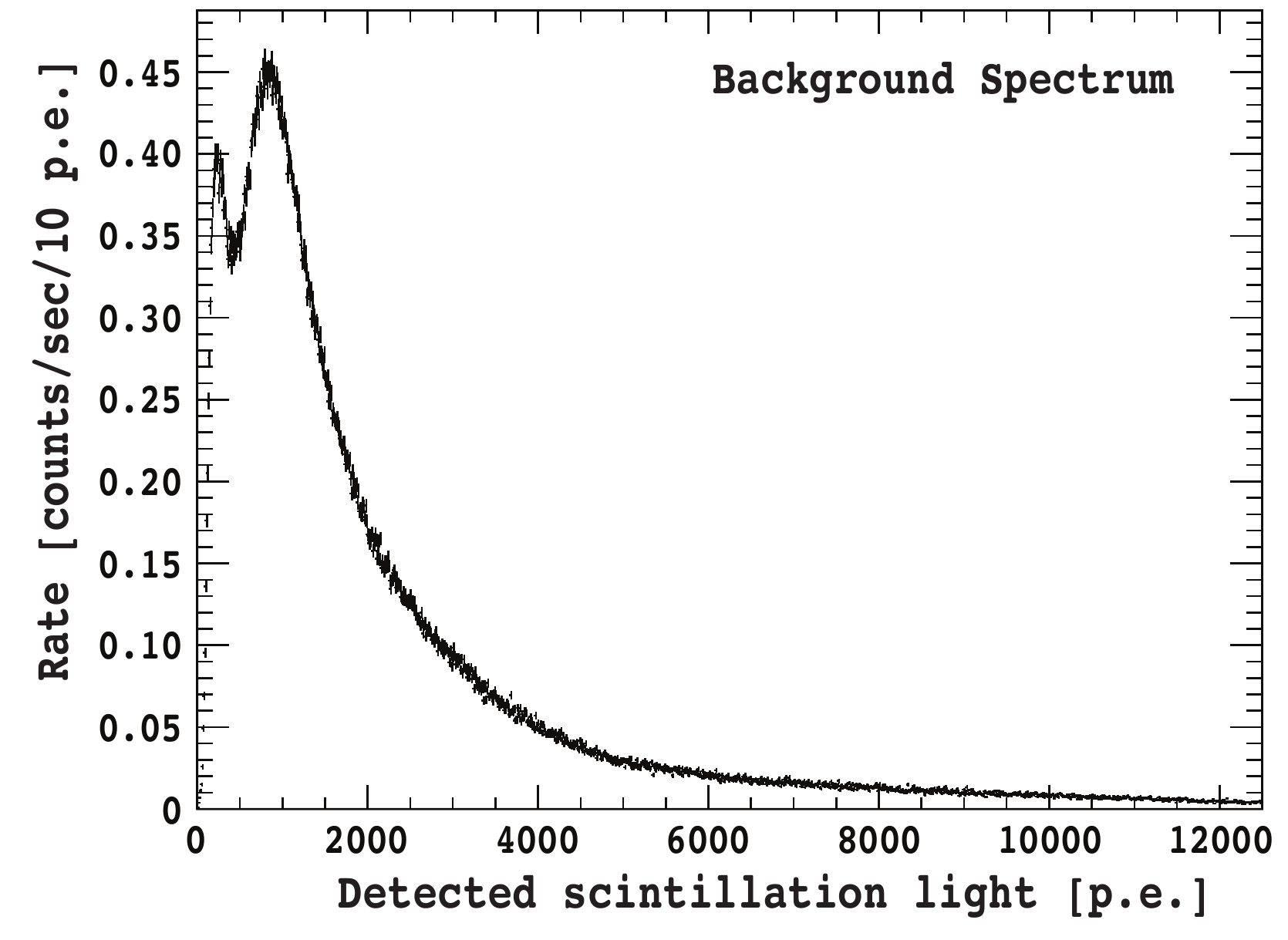}
\caption{Background spectrum acquired with no calibration sources present.}
\label{fig:spectrum_bkgd}
\end{center}
\end{figure}

\indent
Events with known energy depositions are obtained by exposing the detector to a series of external radioactive $\gamma$ sources.  The gamma sources are collimated by means of a 45.5-mm-thick lead collimator with a 10 mm diameter hole. Three collimator positions along the vertical direction are used: \emph{bottom}, \emph{central}, and \emph{top}, corresponding to 20~mm, 105~mm, and 158~mm from the TPC cathode. Due to the large amount of material between the source collimator (located outside the LAr Dewar) and the active volume, as well as the relatively large size of the sensitive region, the spectra are degraded, leaving full-absorption peaks as the most visible and reliable features for light yield estimates.
\begin{table}[t]
   \centering
   \begin{tabular}{|c|c|c|c|}
    \hline \hline
      Source  & E$_{\gamma}$ [keV] & I$_{\gamma}$ \% & Activity [$\mu$Ci] \\ \hline \hline
      $^{57}$Co & 14.41 & 9.16 & \multirow{3}{*}{0.96}\\
      $^{57}$Co & 122.06 & 85.60 &\\
      $^{57}$Co & 136.47 & 10.68 & \\
      \hline
      $^{22}$Na & 510.99 & 180.76& \multirow{2}{*}{1.08}\\
      $^{22}$Na & 1274.53 & 99.94 &\\
      \hline
      $^{137}$Cs & 661.66 & 85.10 & 0.94\\
      \hline 
         \end{tabular}
   \caption{$\gamma$ energies and intensities~\cite{Lbnl:2012:ti}, and activities of sources used.  The 511-keV $^{22}$Na line results from positron annihilation resulting in two back-to-back 511-keV $\gamma$ rays. }
   \label{tab:isotope}
\end{table}
\indent
Light yield measurements have been performed with $^{57}$Co, $^{22}$Na, and $^{137}$Cs whose main gamma energies and intensities are summarized in Table~\ref{tab:isotope}. The data for a single spectrum contain about 1,000,000 events taken over a few hours.  Gamma rays interact in the active volume through  Compton scattering and the photoelectric effect, with events in the full-energy peak typically resulting from multiple interactions.    Figures~\ref{fig:spectrum_co},~\ref{fig:spectrum_cs}, and~\ref{fig:spectrum_na} show the gamma-induced scintillation spectra obtained with the three sources collimated at the central position, after subtraction of a background spectrum (Fig.~\ref{fig:spectrum_bkgd}) acquired with no source present and scaled by the ratio of the livetimes. The events in the plots were analyzed using the linear-baseline algorithm. 

A minimal set of cuts is applied in order to remove from the spectra:
\begin{itemize}  
  \setlength{\itemsep}{0pt}
  \setlength{\parskip}{0pt}
  \setlength{\parsep}{0pt}
\item events saturating the digitizer ADC of any channel;
\item events with a rejected baseline (described above);
\item events in which the first found pulse in the acquisition window is not within 100 ns of the trigger time.
\end{itemize}
These cuts typically retain $>\!\!98\%$ of the triggered events.

\begin{table}[t]
   \centering
   \begin{tabular}{|c|c|c|c|}
    \hline \hline
     E$_{\gamma}$ & $\mu_p$ & $\sigma_p$ & LY$_{\gamma}$ \\ 
     {[keV$_{ee}$]} &  [p.e.] & [p.e.] & [p.e./keV$_{ee}$]\\ \hline \hline
      122.06 & 1082.0$\pm$2.3 &  56.80 & 8.865$\pm$0.019 \\
       510.99 & 4486.4$\pm$2.5 & 152.86 & 8.780$\pm$0.007\\
       661.657 & 6009.6$\pm$1.8 &  186.19 & 9.083$\pm$0.005\\
     1274.53 & 10961.9$\pm$6.7 & 318.07 & 8.601$\pm$0.007\\  \hline
   \end{tabular}
   \caption{Fitted gamma full-absorption peak mean, width, and light yield.  The error on $\mu_p$ is the statistical error from the fit.  The error on LY$_{\gamma}$ is the fit error combined with the statistical error on the mean single-p.e.~response.}
   \label{tab:fits}
\end{table}

Looking at the scintillation spectra in Figs.~\ref{fig:spectrum_co}-\ref{fig:spectrum_na} it is evident that, while the Compton edges are degraded, the full-absorption peaks are very clear.  This is consistent with expectations from a GEANT4-based Monte Carlo simulation of the experimental setup, including the material between the source and the active volume. The experimental spectra in the region of the full absorption peaks are fitted with a Gaussian function with mean value $\mu_p$ and standard deviation $\sigma_p$. For $^{22}$Na and $^{137}$Cs, where the Compton edge and degraded gamma tails are more significant, a falling exponential term is added to the fit function. For $^{57}$Co the full absorption peaks of the 122 and 136 keV lines are not individually resolved and hence the spectrum was fit with the sum of two Gaussians. The ratio between the means and variances of the two Gaussians were fixed to the ratio of the energies and the ratios of the integrals were fixed to the relative intensity of the two $\gamma$ rays (see Table \ref{tab:isotope}). Fitting these functions to simulated spectra reproduces the true peak positions to better than 1\%. The best-fit functions are also shown in Figs.~\ref{fig:spectrum_co}-\ref{fig:spectrum_na}.  The best-fit values of the parameters of interest are summarized in Table~\ref{tab:fits} together with the light yield (LY$_{\gamma}$), defined for each fit as $\mu_p$/E$_{\gamma}$.

\indent
The measured widths of the full absorption peaks deserve a separate discussion. As reference we obtain an energy resolution of 3.1\% ($\sigma_p/\mu_p$) for 662 keV $\gamma$ rays. The variance of the detector response, in photoelectrons, to a mono-energetic energy release can be described as~\cite{Saldanha:2012rs}
\begin{equation}
  \sigma_p^2 = \sigma^2_\text{baseline} + \sigma^2_\text{pe} + \sigma^2_\text{PMT} + \sigma^2_\text{geom}
   \label{eq:sigma}
\end{equation}
where
\begin{itemize}
  \setlength{\itemsep}{0pt}
  \setlength{\parskip}{0pt}
  \setlength{\parsep}{0pt}
\item $\sigma_\text{baseline}^2$ is the variance of the baseline noise, integrated over the length of the pulse. It increases with the amplitude of the scintillation signals and, for 662 keV $\gamma$ rays, has a typical value of 2800 p.e.$^2$;
\item $\sigma_\text{pe}^2$ is the variance in the number of photoelectrons produced. We assume the scintillation process to be Poissonian, with the variance equal to the mean number of photoelectrons, $\mu_p$; 
\item $\sigma^2_\text{PMT}$ is the variance of the photomultiplier response. For PMTs with a similar response, it is approximated by $\sigma^2_{\psi_1}\cdot \mu_p$, where $\sigma^2_{\psi_1}$ is the relative variance of the single-photoelectron response (see Eq.~\ref{eq:pdf}) averaged over all channels, with a typical value of 0.2;
\item $\sigma^2_\text{geom}$ is the geometrical variance, associated with spatial non-uniformities in the light collection of the detector. Due to their different mean interaction lengths in liquid argon, gammas of different energies can probe different regions of the active volume. Thus, the geometrical variance is expected to be different for different sources. The total variance due to geometrical effects can be written as $\sigma^2_\text{LY}\cdot{\mu_p}^2$, where $\sigma^2_\text{LY}$ is the relative variance of the light yield over the interaction region. The order of magnitude of this term can be qualitatively compared to the observed variation of LY$_\gamma$ with the vertical position of the source. $^{22}$Na source runs performed with the collimator located in the top and bottom positions show, respectively, a decrease of 4.4\% and an increase of 5.6\% in the light yield with respect to the central position. This asymmetry agrees with expectations as the vertical symmetry of the system is broken by the presence of the liquid-gas interface and the grid near the top, both of which favor light collection by the bottom PMT's. We note that in full TPC mode, three-dimensional event reconstruction allows these non-uniformities in the detector response to be measured and corrections applied. 
\end{itemize}
From the estimates for the individual terms above, at 662 keV we obtain an energy resolution of $\sim$ 0.9\% from the baseline, $\sim 1.3\%$ from photoelectron statistics, and $\sim$ 0.6\% from the PMT response. The residual resolution in the observed response ($\sim2.6$\%) is of the same order of magnitude as the estimated contribution for the geometrical variance. It should be noted that Eq.~\ref{eq:sigma} does not account for any additional variance from multiple Compton scattering (such as non-linear quenching) or possible non-Poissonian fluctuations in the distribution of scintillation photons \cite{Doke1976353, 2003PhRvB..68e4201C, PhysRevB.76.014115}.   
   
\begin{figure}[t]
\begin{center}
\includegraphics[width = \linewidth]{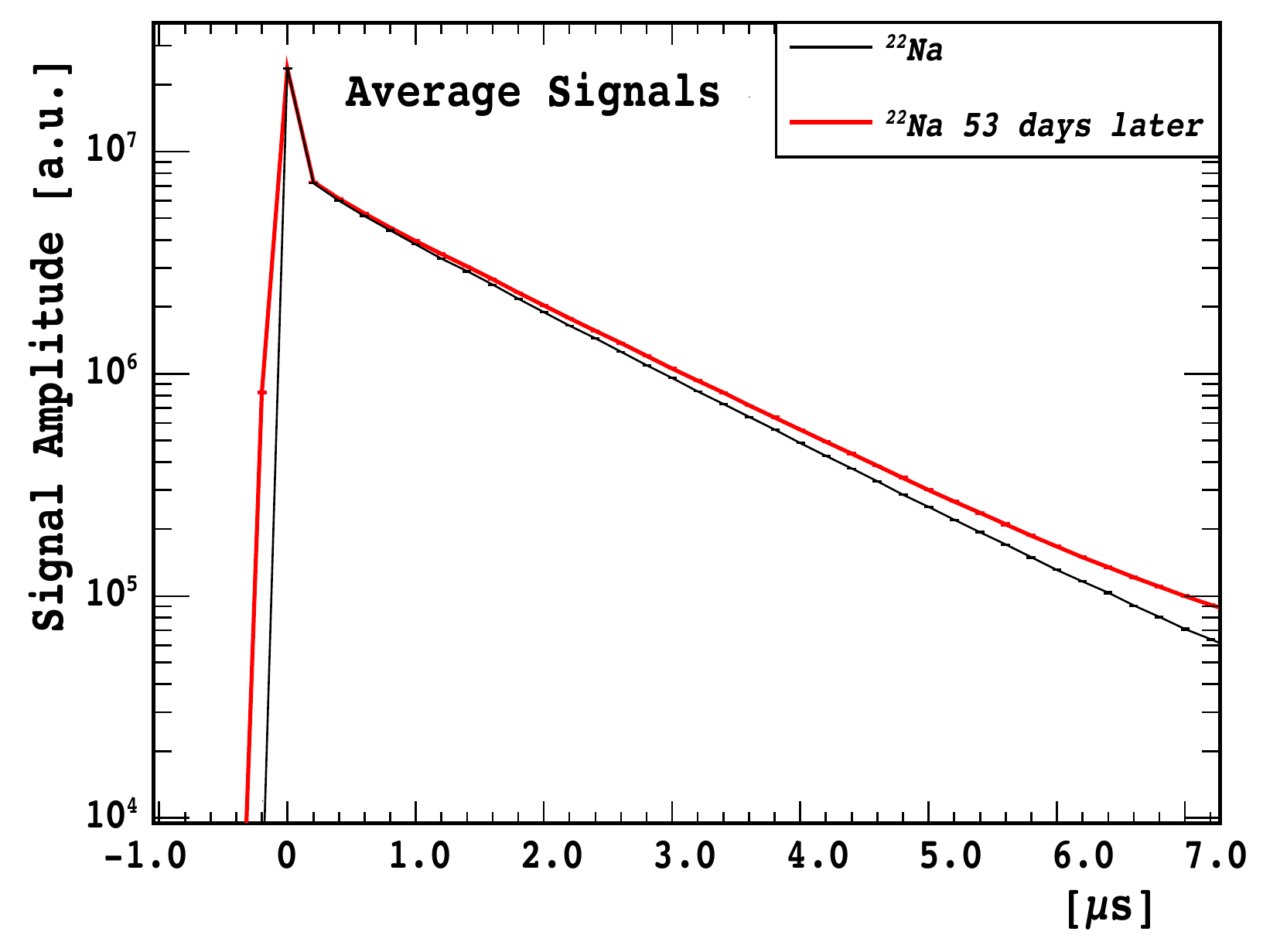}
\caption{Average scintillation waveforms from a single PMT in 0.2~$\mu$s bins.  The waveforms are from two $^{22}$Na run collimated at the central position, one (red) taken 53~days after the other (black). 
There has been a clear increase in the slow-component lifetime between the runs.
The change in the leading edge at the left of the plot is thought to be due to a small change in the trigger timing.}
\label{fig:lifetime}
\end{center}
\end{figure}

Light collection performance has shown good stability with time. A $^{22}$Na calibration run (collimated at the central position) performed 53~days after the one shown in Table~\ref{tab:fits} gives LY$_\gamma=9.142\pm$0.006~p.e./keV$_{ee}$ for the 511~keV line. The observed light yield increase of about 4\% is likely associated with an improvement in the liquid argon purity due to the  running of the purification system between the two measurements. Argon contaminants such as N$_{2}$ and O$_{2}$ are known to quench the argon scintillation light via non-radiative collisional de-excitation~\cite{2010JInst...5.5003A, 2010JInst...5.6003A}.  This process also reduces the observed slow-component lifetime.
Figure~\ref{fig:lifetime} shows average scintillation waveforms from the two runs. Independent of any particular model, the slow-component lifetime has clearly improved from the first to the second run, suggesting the elimination of de-exciting contaminants. The fit to an exponential in the range 1.0-5.0~$\mu$s provides lifetimes of (1.4601$\pm$0.0007)~$\mu$s for the first run and (1.5349$\pm$0.0008)~$\mu$s for the second,
where the errors are statistical only. A simple model with an absolute-purity slow-component lifetime of 1.6~$\mu$s, predicts that this increase in lifetime would correspond to an increase in total light yield of 3.8\%, in good agreement with that observed.

Several sources of systematic uncertainty have been considered and are summarized in Table~\ref{tab:fit_syserr}. As discussed in Sec.~\ref{sec:source}, the algorithm used to evaluate the baseline affects the integral of the digitized signals. A study of the effect of the baseline algorithm on simulated data has shown that the moving-baseline algorithm tends to underestimate the true integral for events with a large number of photoelectrons. Nonetheless, we include the difference in $^{137}$Cs light yields between the two baseline algorithms as a systematic
uncertainty in Table~\ref{tab:fit_syserr}.

A second source of systematic uncertainty is the function modeling the spectrum. One component of this uncertainty is the use of an exponential to model the spectrum under the Gaussian in the full-absorption-peak fits. We conservatively estimate this uncertainty by re-fitting the $^{137}$Cs peak with a Gaussian only.  The observed variation in the fit result is 0.07\%. A contribution of the same order is attributed to the background subtraction, estimated by re-fitting the $^{137}$Cs spectrum without subtracting the background. Fitting simulated $^{137}$Cs and $^{22}$Na spectra with the same Gaussian+exponential used on data shows systematic displacement of the fitted peak from the true value, typically 0.7\%.  We combine these three components into the ``Fit function" entry in Table~\ref{tab:fit_syserr}.

In the fit of the single-photoelectron spectrum, the parameters of the exponential term have shown some instability when noise increases the pedestal width.  This can result in sizable excursions in individual channels.  To explore this, we measured the values of the exponential parameters for each PMT using a single laser run, chosen to be relatively clean.  The full laser calibration was redone with these parameters fixed and the calibration was used to reanalyze the source spectra.  Shifts of up to 0.5\% are observed in the resulting light yields and we assign this as a systematic error. We vary the spectrum binning and the integration region to estimate systematic uncertainties associated with the mechanics of the single-photoelectron fit.  These variations have $\sim\!\!1\%$ effects on the light yield estimate and, combined with the systematic uncertainty from the exponential term, are included in Table~\ref{tab:fit_syserr}.
\begin{table}[t]
   \centering
   \begin{tabular}{|c|c|}
    \hline \hline
      Source & [\%] \\ \hline \hline
      Baseline algorithm &  4.9 \\
      Fit function & 0.7\\
      Single photoelectron & 1.0\\ \hline
      Total & 5.0\\
    \hline \hline
   \end{tabular}
   \caption{Systematic uncertainties in light yield measurement [\%]}
   \label{tab:fit_syserr}
\end{table}

\section{Conclusions}
\indent
The light yield reported here greatly exceeds that measured in the previous run of DS-10, about 4.5~p.e./keV$_{ee}$. Since the previous run, a number of modifications were made to the detector.  The most relevant of these were the replacement of the bottom PMT array and the replacement of the top and bottom windows.  In the previous run, the bottom PMT array consisted of a single 8" Hamamatsu R5912-02 PMT.  This was replaced with an array of 7 3" R11065s, matching the top array.  The new array had less photocathode coverage (partly compensated by filling the gaps between the 3" PMTs with PTFE reflectors), but much higher quantum efficiency (averaging 33.9\% vs.~18\%).  In the previous run, the windows were acrylic, with 100-nm-thick ITO on  both sides.  A 100 nm ITO layer is calculated to absorb 10\% of 420 nm light at normal incidence, with a large effect on the light yield when multiple passes and non-normal incidence were considered.  However, thinner coatings were not recommended on acrylic. The replacement windows, fused silica with 15-nm ITO, were expected to provide  considerably better light yield.  

As described in Sec.~\ref{sec:spe} the light yields reported here depend directly on the single-photoelectron calibration, in which the response was fitted to the single-p.e. model of Eq.~\ref{eq:pdf}.  The first, exponential, term lowers the single-p.e.~mean, and thus raises the inferred number of p.e. in a signal of a given integrated charge. The presence of such a term in the single-p.e. response of the PMT's is motivated by  structure below the single-p.e. peak observed to be correlated with laser activity.   However, its inclusion in the light yield may not be appropriate for all applications, notably  those that count single photoelectrons above some threshold.  As discussed in  Sec.~\ref{sec:spe}, including less of the exponential term is at most a 13\% effect.

The light yield achieved in DarkSide-10, 9.142$\pm$0.006(stat)$\pm$0.457(sys)~p.e./keV$_{ee}$ after  the purification campaign, is well in excess of the light yields proposed in Reference~\cite{Boulay2006179} and assumed in the background calculations for the 50-kg DarkSide-50~\cite{darkside}.  It demonstrates that excellent light yield can be achieved in the elaborate structure of a TPC.
\\ \\
We acknowledge support from the NSF (U.S., Grants PHY-0919363, PHY-1004072, and associated collaborative Grants), DOE (U.S., Contract Nos. DE-FG02-91ER40671 and DE-AC02-07CH11359), and the Istituto Nazionale di Fisica Nucleare (Italy).

\bibliographystyle{model1-num-names.bst}

\end{document}